\documentstyle[epsf,epsfig]{prhep97}

\def\gsimfig{\mathrel{\raise.2ex\hbox{$>$}\hskip-.8em\lower.9ex\hbox{$\sim$}}}

\def\lsimfig{\mathrel{\raise.2ex\hbox{$<$}\hskip-.8em\lower.9ex\hbox{$\sim$}}}
\def\gsimfig{\mathrel{\raise.2ex\hbox{$>$}\hskip-.8em\lower.9ex\hbox{$\sim$}}}

\def\mboxsc#1{\mbox{\scriptsize #1}}

\def\docuname{{\large \sc  mepjet}}

\def\ap#1#2#3   {{\em Ann. Phys. (NY)} {\bf#1} (#2) #3}
\def\apj#1#2#3  {{\em Astrophys. J.} {\bf#1} (#2) #3}
\def\apjl#1#2#3 {{\em Astrophys. J. Lett.} {\bf#1} (#2) #3}
\def\app#1#2#3  {{\em Acta. Phys. Pol.} {\bf#1} (#2) #3}
\def\ar#1#2#3   {{\em Ann. Rev. Nucl. Part. Sci.} {\bf#1} (#2) #3}
\def\cpc#1#2#3  {{\em Computer Phys. Comm.} {\bf#1} (#2) #3}
\def\err#1#2#3  {{\it Erratum} {\bf#1} (#2) #3}
\def\ib#1#2#3   {{\it ibid.} {\bf#1} (#2) #3}
\def\jmp#1#2#3  {{\em J. Math. Phys.} {\bf#1} (#2) #3}
\def\ijmp#1#2#3 {{\em Int. J. Mod. Phys.} {\bf#1} (#2) #3}
\def\jetp#1#2#3 {{\em JETP Lett.} {\bf#1} (#2) #3}
\def\jpg#1#2#3  {{\em J. Phys. G.} {\bf#1} (#2) #3}
\def\mpl#1#2#3  {{\em Mod. Phys. Lett.} {\bf#1} (#2) #3}
\def\nat#1#2#3  {{\em Nature (London)} {\bf#1} (#2) #3}
\def\nc#1#2#3   {{\em Nuovo Cim.} {\bf#1} (#2) #3}
\def\nim#1#2#3  {{\em Nucl. Instr. Meth.} {\bf#1} (#2) #3}
\def\np#1#2#3   {{\em Nucl. Phys.} {\bf#1} (#2) #3}
\def\pcps#1#2#3 {{\em Proc. Cam. Phil. Soc.} {\bf#1} (#2) #3}
\def\pl#1#2#3   {{\em Phys. Lett.} {\bf#1} (#2) #3}
\def\prep#1#2#3 {{\em Phys. Rep.} {\bf#1} (#2) #3}
\def\prev#1#2#3 {{\em Phys. Rev.} {\bf#1} (#2) #3}
\def\prl#1#2#3  {{\em Phys. Rev. Lett.} {\bf#1} (#2) #3}
\def\prs#1#2#3  {{\em Proc. Roy. Soc.} {\bf#1} (#2) #3}
\def\ptp#1#2#3  {{\em Prog. Th. Phys.} {\bf#1} (#2) #3}
\def\ps#1#2#3   {{\em Physica Scripta} {\bf#1} (#2) #3}
\def\rmp#1#2#3  {{\em Rev. Mod. Phys.} {\bf#1} (#2) #3}
\def\rpp#1#2#3  {{\em Rep. Prog. Phys.} {\bf#1} (#2) #3}
\def\sjnp#1#2#3 {{\em Sov. J. Nucl. Phys.} {\bf#1} (#2) #3}
\def\spj#1#2#3  {{\em Sov. Phys. JEPT} {\bf#1} (#2) #3}
\def\spu#1#2#3  {{\em Sov. Phys.-Usp.} {\bf#1} (#2) #3}
\def\zp#1#2#3   {{\em Zeit. Phys.} {\bf#1} (#2) #3}
\makeatletter
\let\chapter\hid@chapter
\makeatother
\begin{document}
%\pagenumbering{empty}

% The following definitions need to be customised;

% Will appear on page headings
\authorrunning{E.~Mirkes, S.~Willfahrt and D.~Zeppenfeld}
\titlerunning{{\talknumber}: Jet Production in DIS with
                             $Z$ and $W$ Exchange}
 
% Now the full name of author and talk

% For plenary talks, the talk number is that of the session
\def\talknumber{406} 

\title{{\talknumber}:
        Jet Production in DIS at NLO \\Including $Z$ and $W$ Exchange}
\author{\underline{E.~Mirkes$^a$}, S.~Willfahrt$^a$ and D.~Zeppenfeld$^b$}
\institute{$^a$Institut f\"ur Theoretische Teilchenphysik, 
           Universit\"at Karlsruhe,\\ D-76128 Karlsruhe, Germany\\
           $^b$Department of Physics, 
           University of Wisconsin, Madison, WI 53706, USA\\[-15mm]}
\maketitle
\vspace{-6cm}
\hfill \vtop{   \hbox{\bf hep-ph/9711366}
                \hbox{\bf TTP97-47}}\footnote{Talk given 
by E. Mirkes at the International Europhysics Conference on High-Energy Physics
(HEP 97), Jerusalem, Israel, 19-26 Aug 1997. }
\vspace{6cm}

\begin{abstract}
Next-to-leading order QCD predictions for 1-jet and 2-jet cross sections in
deep inelastic scattering with complete neutral current 
($\gamma^*$ and $Z$) and charged current
($W^\pm$) exchange are presented.
\end{abstract}
\begin{figure}[ht]
  \centering
  \mbox{\epsfig{file=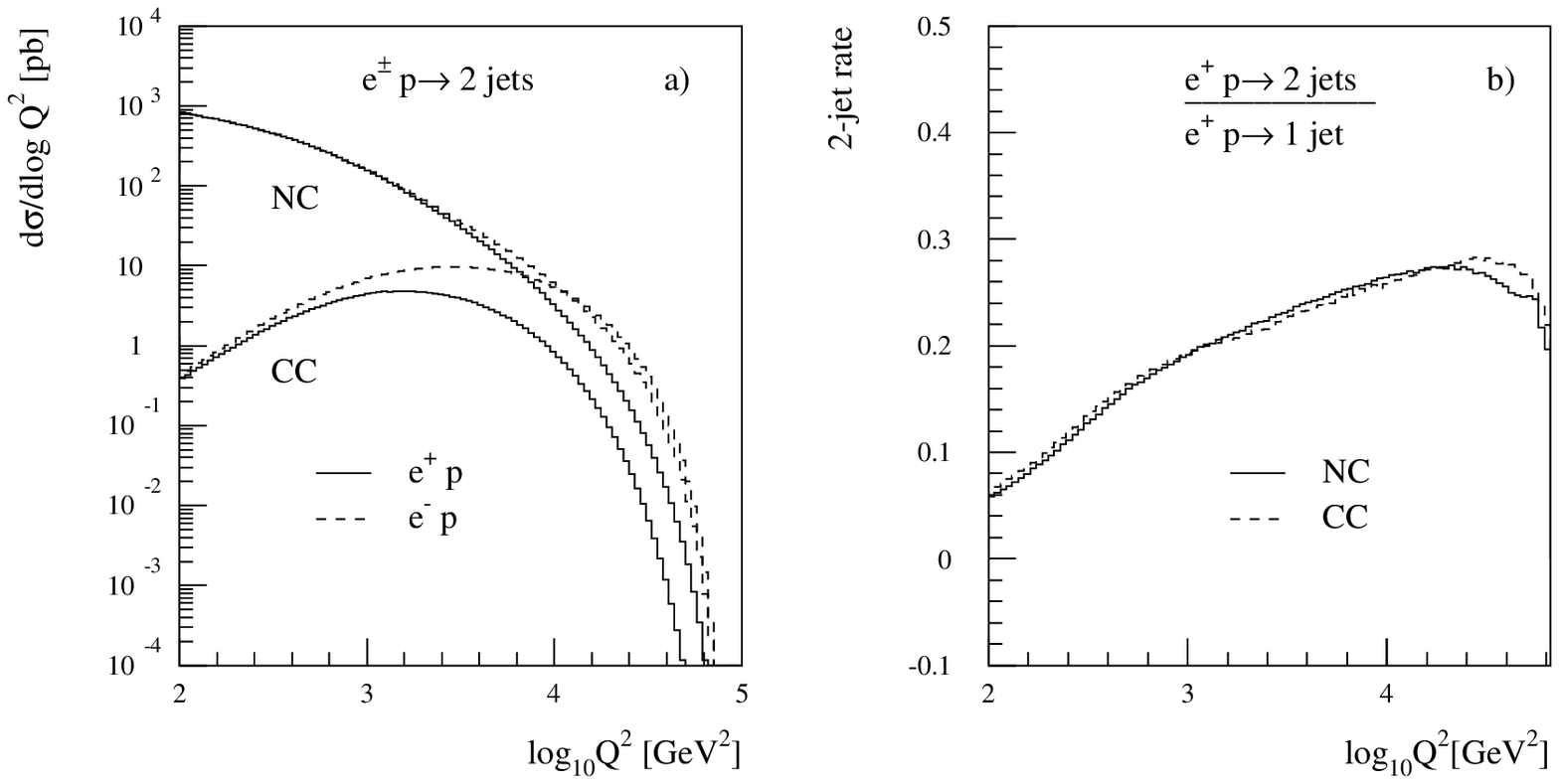,bbllx=0,bblly=250,
               bburx=550,bbury=550,width=0.95\linewidth}}\\[-8mm]
  \mbox{\epsfig{file=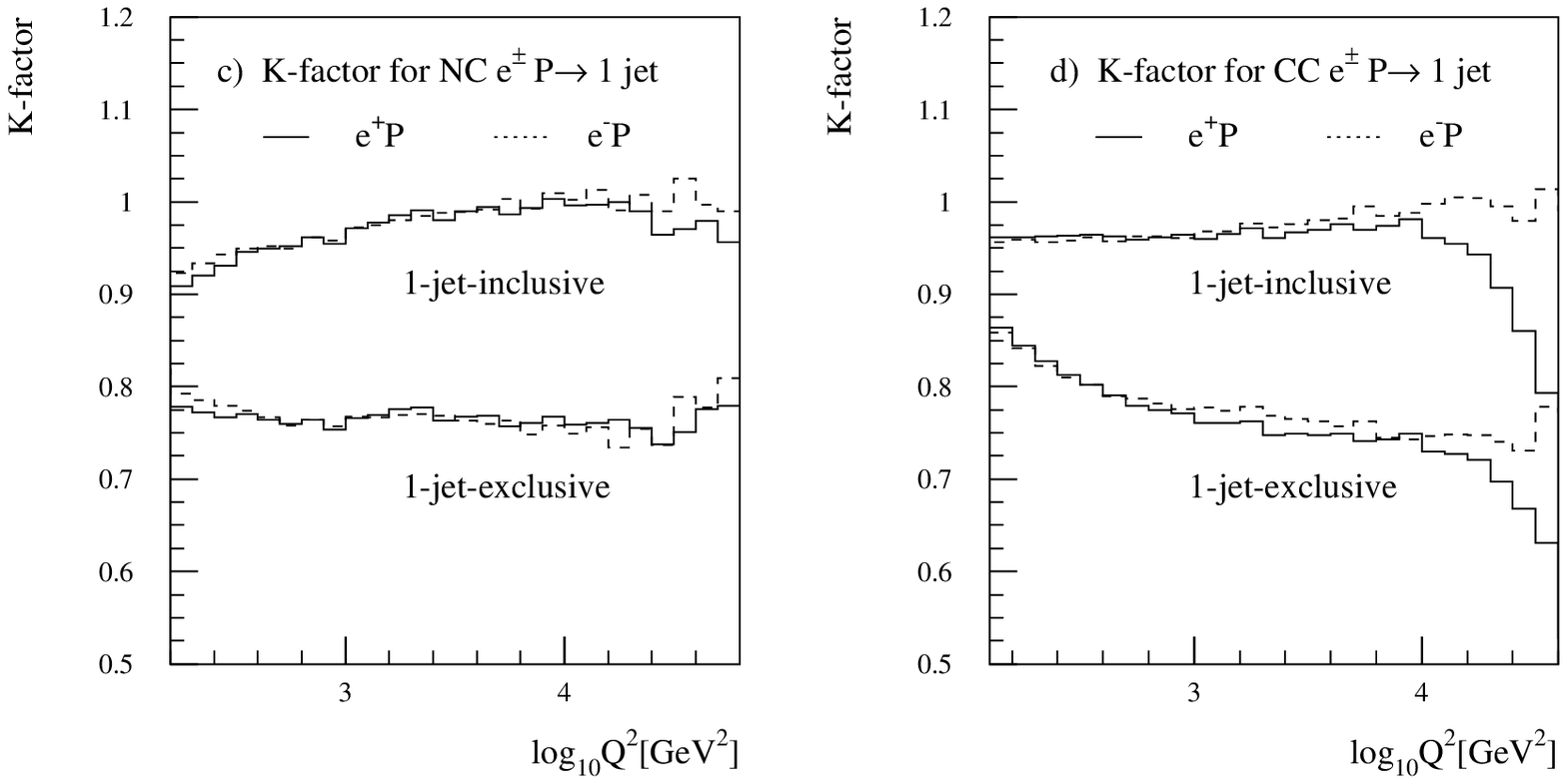,
         bbllx=0,bblly=240,bburx=540,bbury=540,
         width=0.95\linewidth}}\\[-8mm]
  \mbox{\epsfig{file=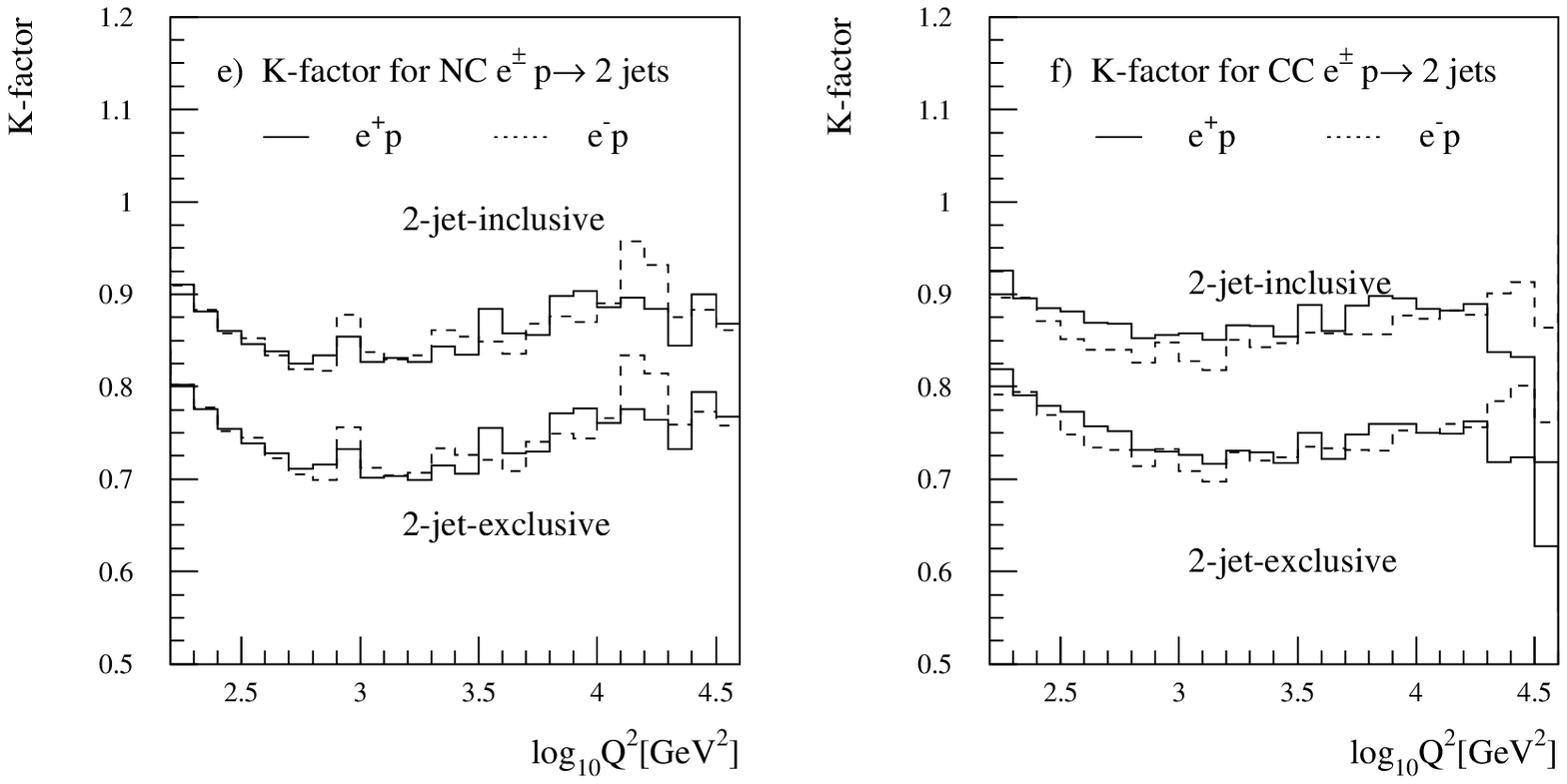,
         bbllx=0,bblly=240,bburx=540,bbury=540,
         width=0.95\linewidth}} 
\vspace*{-8mm}
\caption{
a) $Q^2$ dependence of the  CC 
and NC 2-jet cross section for
$e^+p$ (solid) and $e^-p$ (dashed) scattering.
b) 2-jet rate $\sigma(\protect\mbox{2-jet})[Q^2]/
               \sigma(\protect\mbox{1-jet})[Q^2]$ for NC (solid) and CC 
(dashed) exchange. The results in a,b are shown in LO with
MRS Set (R1) \protect\cite{mrsr1}
parton distributions and the two-loop formula
for the strong coupling constant.\protect\newline
c,d)
$Q^2$ dependence of the $K$-factor for NC 
and CC  1-jet inclusive and exclusive
cross sections in $e^+p$ (solid) and $e^-p$ (dashed) scattering.
The (LO) NLO  cross section, which enter the $K$-factor,
are calculated with LO (NLO) GRV parton distributions 
\protect\cite{grv} together
with the 1-loop (2-loop) formula for the strong coupling constant.
e,f)  same as c,d) for 2-jet cross sections.
}
\label{fig}
\end{figure}

\enlargethispage{1cm}
Electroweak effects in DIS $n$-jet production 
are known to become important for $Q^2\gsimfig 2500$ GeV$^2$.
These effects can be studied with the
fully differential $ep \rightarrow n$ jets event generator \docuname\ 
\cite{plb1,habil}
which allows to analyze any 1- or 2-jet  like
observable in NLO in terms of parton 4-momenta.
Numerical results for 1- and 2-jet cross sections in $e^\pm p$ scattering
are shown in Fig.~\ref{fig}.

Jets are defined in a cone scheme 
(in the laboratory frame) with  $R=1$
and $p_T^{\mboxsc{lab}}(j)>5$~GeV.
In addition, we require
$0.04 < y < 1$, an energy cut of $E(e^\prime)>10$~GeV on the scattered 
lepton (in NC scattering) and a cut on the pseudo-rapidity 
$\eta$
%$
%\eta=-\ln\tan(\theta/2)
%$
of the scattered lepton and jets of $|\eta|<3.5$. 

The $Q^2$ distribution of the NC 2-jet cross section (including
$\gamma^\ast$ and $Z$ exchange)
is compared to the CC  2-jet cross section in Fig.~\ref{fig}a.
The difference between the
$e^+p$ and $e^-p$ results in the NC case is entirely
due to the additional $Z$ exchange, which increases (decreases)
the 1-photon result in the $e^-p$
($e^+p$) case by more than a factor of two for very high $Q^2$.
The $Q^2$-dependence of the 
2-jet rate, defined as the ratio of the 2-jet to the 1-jet
cross section, is shown in Fig.~\ref{fig}b
for $e^+p$ scattering in NC (solid) and CC (dashed) exchange.

The QCD corrections to the electroweak cross sections are investigated in
Figs.~\ref{fig}c-f where the $Q^2$ dependence of the
$K$-factors for NC and CC  1- and 2-jet cross sections are shown.
While NC and CC effects  differ markedly for $e^+p$ and
$e^-p$ scattering,  the QCD $K$-factors are largely
independent of the initial state lepton.
We find very similar $K$-factors for jets defined in the $k_T$
scheme with $E_T^2=40$ GeV$^2$.
More results can be found in \cite{habil}.

% ---- Bibliography ----
%


\begin{thebibliography}{99}

\bibitem{plb1}
E. Mirkes and D. Zeppenfeld, \pl{B380}{1996}{105},\,  [hep-ph/9511448];
{\em Acta Phys. Pol.} {\bf B27} (1996) 1393,\, [hep-ph/9604281].


\bibitem{habil}
E. Mirkes, [hep-ph/9711224]. 

\bibitem{grv}
M. Gl\"uck, E. Reya and A. Vogt, \zp{C67}{1995}{433}.

\bibitem{mrsr1}  
A.D.~Martin, R.G.~Roberts, W.J.~Stirling, \pl{B387}{1996}{419}.

\end{thebibliography}
\end{document}